\documentclass{amsart}
\usepackage{amssymb}
\usepackage{amsmath}
\usepackage{amsfonts}

\newtheorem{theorem}{Theorem}
\theoremstyle{plain}

\newtheorem{corollary}{Corollary}

\newtheorem{proposition}{Proposition}
\newtheorem{remark}{Remark}

\numberwithin{equation}{section}
\input{tcilatex}

\begin{document}
\title[Inverse scattering method for nonlinear Klein--Gordon equation]{%
Inverse scattering method via Gel'fand--Levitan--Marchenko equation for some
negative order nonlinear wave equations}
\author{Mansur I. Ismailov$^{\ast ),\text{ }\ast \ast )}$ and Cihan Sabaz$%
^{\ast )}$ }
\email{mismailov@gtu.edu.tr; cihansabaz@gmail.com}
\date{}
\dedicatory{$^{\ast )}$\textit{\ Department of Mathematics, Gebze Technical
University, \ 41400 Gebze-Kocaeli, Turkey}\\
$^{\ast \ast )}$ \textit{Department of Mathematics, Khazar University, 1096
Baku, Azerbaijan}}

\begin{abstract}
A class of negative order Ablowitz--Kaup--Newell--Segur nonlinear evolution
equations are obtained by applying the Lax hierarchy of the first order
linear system of three equations. The inverse scattering problem on the
whole axis are examined in the case where linear system becomes the
classical Zakharov--Shabat system consists of two equations and admits a
real symmetric and real anti-symmetric potential. Referring to these
results, the N--soliton solutions for the integro-differential version of
the nonlinear Klein--Gordon equation coupled with a scalar field (CKG) and
negative order modified Korteweg-de Vries (nmKdV) equation are obtained by
using the inverse scattering method via the Gel'fand--Levitan--Marchenko
equation.
\end{abstract}

\maketitle

\section{Introduction}

A completely integrable nonlinear equation of mathematical physics is one
which has a Lax representation, or, more precisely, can be solved via a
linear integral equation of Gel'fand--Levitan--Marchenko (GLM) type, the
classic examples being the Korteweg--de Vries, sine-Gordon and nonlinear Schr%
\"{o}dinger equations, $[1,2,3]$. It is applied the AKNS hierarchy to derive
soliton solutions of these integrable models\ by the inverse scattering
method. Recently, many integrable hierarchies of soliton equations have been
extended to hierarchies of a negative order Ablowitz--Kaup--Newell--Segur
(AKNS) equation by many authors, $[4,5,6,7]$. This gives an useful necessary
extension for complete integrability, which is applied to investigate the
integrability of certain generalizations of the Klein--Gordon equations,
some model nonlinear wave equations of nonlinear Klein--Gordon equation
coupled with a scalar field.

Consider the nonlinear Klein--Gordon equation coupled with a field $v$, in
the following form, $[8]$:

\begin{equation}
\left\{ 
\begin{array}{c}
u_{\varkappa \varkappa }-u_{\tau \tau }-u+2u^{3}+2vu=0, \\ 
v_{\varkappa }-v_{\tau }-4uu_{\tau }=0.%
\end{array}%
\right.  \tag{1.1}
\end{equation}

In the case $v\neq 0$, this equation is integrable since it admits the same
bilinear form with the well-known sine-Gordon equation, $[8]$.

The coupled nonlinear Klein--Gordon equations are analyzed for their
integrability properties in $[9]$ where the Hirota bilinear form is
identified, from which one-soliton solutions are derived. Then, the results
are generalized to the two, three and N-coupled Klein--Gordon equations in $%
[10,11]$. Another direct method for traveling wave solutions of coupled
nonlinear Klein--Gordon equations is employed in $[12]$.

\bigskip The equation (1.1) becomes the following negative first order
equation:

\begin{equation}
r_{tx}=2r\partial _{x}^{-1}\left[ \left( r^{2}\right) _{t}\right] +r 
\tag{1.2$_{+}$}
\end{equation}%
by the change of variables $\varkappa =\frac{x+4t}{2}$, $\tau =\frac{4t-x}{2}
$, elimination of $v$ in second equation under the assumption that the
scalar field $v$ tends to zero at infinity and by the substitution $r=\frac{1%
}{2}u$, where $\partial _{x}^{-1}=\int\limits_{x}^{\infty }dx$\ is
indefinite integral with respect to $x$.

The similar equation to (1.2$_{+}$) with negative sign on integral term 
\begin{equation}
r_{tx}=-2r\partial _{x}^{-1}\left[ \left( r^{2}\right) _{t}\right] +r 
\tag{1.2$_{-}$}
\end{equation}%
is called the negative order modified Korteweg-de Vries (nmKdV) in [13] as
it is in the negative flow of the mKdV hierarchy [14].

Our aim in this paper is to find the soliton solutions of (1.2$_{+}$) and
(1.2$_{-}$) by the inverse scattering method. The inverse scattering method
is the most important discovery in the theory of soliton. It provides us
alternatively show the complete integrability of the nonlinear evolution
equation. This method also enables to solve the initial value problem for
nonlinear evolution equations (1.2$_{+}$) and (1.2$_{-}$). Shortly we call
the equation (1.2$_{+}$) the CKG and (1.2$_{-}$) the nmKdV equations in
future.

The brief outline of the paper is the followings. In Section 2, we find that
the CKG and nmKdV equations possesses a Lax pair of the negative order AKNS
equation. It is shown that the auxiliary systems corresponding to CKG is
classical Zakharov--Shabat (ZS) system with real and anti-symmetric
potential and corresponding to nmKdV is classical Zakharov--Shabat (ZS)
system with real and symmetric potential. Then, in Section 3, we recall the
necessary result on the inverse scattering problem for the ZS and nmKdV
equations on the whole line. In Section 4, we show how the scattering data
evolves when coefficients of ZS system satisfies the CKG and nmKdV
equations. In this section, the N-soliton solutions of the CKG and nmKdV
equations are obtained by inverse scattering method via the GLM
equation.\bigskip

\section{ Negative first Order AKNS Equations}

Consider the spectral problem for 3 $\times $ 3 linear system

\begin{equation}
\left[ 
\begin{array}{c}
\varphi _{1x} \\ 
\varphi _{2x} \\ 
\varphi _{3x}%
\end{array}%
\right] =X(p,q)\left[ 
\begin{array}{c}
\varphi _{1} \\ 
\varphi _{2} \\ 
\varphi _{3}%
\end{array}%
\right] ,  \tag{2.1}
\end{equation}%
where $X(p,q)=i\left[ 
\begin{array}{ccc}
\alpha _{1}\lambda & p_{1} & p_{2} \\ 
q_{1} & \alpha _{2}\lambda & 0 \\ 
q_{2} & 0 & \alpha _{2}\lambda%
\end{array}%
\right] $ with $\lambda $ is a nonzero eigenvalue, $\varphi _{1},\varphi
_{2} $ and $\varphi _{3}$ are linearly independent eigenfunctions, $%
i^{2}=-1, $ $\alpha _{1}$ and $\alpha _{2}$ are real constants with $\alpha
_{1}-\alpha _{2}=\alpha <0$, $p_{1}=p_{1}(x,t)$, $p_{2}=p_{2}(x,t)$, $%
q_{1}=q_{1}(x,t)$ and $q_{2}=q_{2}(x,t)$ are the rapidly decreasing at
infinity complex valued coefficients.

The auxiliary spectral problem described as follows:%
\begin{equation}
\left[ 
\begin{array}{c}
\varphi _{1t} \\ 
\varphi _{2t} \\ 
\varphi _{3t}%
\end{array}%
\right] =T(p,q)\left[ 
\begin{array}{c}
\varphi _{1} \\ 
\varphi _{2} \\ 
\varphi _{3}%
\end{array}%
\right] ,  \tag{2.2}
\end{equation}%
where $T(p,q)=\left[ 
\begin{array}{ccc}
a & b & c \\ 
d & e & f \\ 
k & l & m%
\end{array}%
\right] $ and $a,b,c,d,e,f,k,l$ and $m$ are scalar functions, independent of 
$\varphi _{1},\varphi _{2}$ and $\varphi _{3}.$

\bigskip From (2.1) and (2.2), the zero curvature equation $X_{t}-T_{x}+%
\left[ X,T\right] =0$ yields

\begin{equation*}
\begin{array}{c}
a_{x}=ip_{1}d+ip_{2}k-iq_{1}b-iq_{2}c, \\ 
b_{x}=i\alpha \lambda b-ip_{1}\alpha +ip_{1}e+ip_{2}l+ip_{1t}, \\ 
c_{x}=i\alpha \lambda c+ip_{1}f-ip_{2}\alpha +ip_{2}m+ip_{2t},%
\end{array}%
\begin{array}{c}
f_{x}=iq_{1}c-ip_{2}d, \\ 
l_{x}=iq_{2}b-ip_{1}k, \\ 
m_{x}=iq_{2}c-ip_{2}k,%
\end{array}%
\end{equation*}%
\begin{equation}
\tag{2.3}
\end{equation}%
\begin{equation*}
\begin{array}{c}
d_{x}=-i\alpha \lambda d+iq_{1}a-iq_{1}e-iq_{2}f+iq_{1t}, \\ 
k_{x}=-i\alpha \lambda k-iq_{1}l+iq_{2}a-iq_{2}m+iq_{2t}, \\ 
e_{x}=iq_{1}b-ip_{1}d,%
\end{array}%
\end{equation*}%
where $\alpha =\alpha _{1}-\alpha _{2}.$ Let the following transformations
be applied to the system (2.3):

\begin{eqnarray*}
a &=&\frac{A(x,t)}{\lambda },b=\frac{B(x,t)}{\lambda },c=\frac{C(x,t)}{%
\lambda }, \\
d &=&\frac{D(x,t)}{\lambda },e=\frac{E(x,t)}{\lambda },f=\frac{F(x,t)}{%
\lambda }, \\
k &=&\frac{K(x,t)}{\lambda },l=\frac{L(x,t)}{\lambda },m=\frac{M(x,t)}{%
\lambda }.
\end{eqnarray*}%
As a result the following equations are obtained:

\begin{equation*}
\begin{array}{cc}
A_{x}=ip_{1}D+ip_{2}K-iq_{1}B-iq_{2}C, & E_{x}=iq_{1}B-ip_{1}D, \\ 
B_{x}=ip_{1}E-ip_{1}A+ip_{2}L,\text{ }B=-\frac{1}{\alpha }p_{1t}, & 
L_{x}=iq_{2}B-ip_{1}K, \\ 
C_{x}=ip_{1}F+ip_{2}M-ip_{2}A,\text{ }C=-\frac{1}{\alpha }p_{2t}, & 
M_{x}=iq_{2}C-ip_{2}K,%
\end{array}%
\end{equation*}%
\begin{equation}
\tag{2.4}
\end{equation}%
\begin{equation*}
\begin{array}{c}
D_{x}=iq_{1}A-iq_{1}E-iq_{2}F,\text{ }D=\frac{1}{\alpha }q_{1t}, \\ 
K_{x}=-iq_{1}L+iq_{2}A-iq_{2}M,\text{ }K=\frac{1}{\alpha }q_{2t}, \\ 
F_{x}=iq_{1}C-ip_{2}D,%
\end{array}%
\end{equation*}

The following negative first order AKNS equations are obtained for important
cases of spectral problem (2.1).

\begin{proposition}
If the coefficients of the system (2.1) satisfies the properties $%
q_{1}=-p_{1}$ and $q_{2}=-p_{2}$ then the the compatibility condition (2.4)
becomes the following negative order pair of equations:%
\begin{eqnarray}
p_{1tx} &=&-p_{1}\left[ 2\partial _{x}^{-1}\left( p_{1}^{2}\right)
_{t}+\partial _{x}^{-1}\left( p_{2}^{2}\right) _{t}+c_{1}\right] -p_{2}\left[
\partial _{x}^{-1}\left( p_{1}p_{2}\right) _{t}+c_{2}\right] ,  \notag \\
&&  \TCItag{2.5} \\
p_{2tx} &=&-p_{2}\left[ \partial _{x}^{-1}\left( p_{1}^{2}\right)
_{t}+2\partial _{x}^{-1}\left( p_{2}^{2}\right) _{t}+c_{3}\right] -p_{1}%
\left[ \partial _{x}^{-1}\left( p_{1}p_{2}\right) _{t}+c_{4}\right] ,  \notag
\end{eqnarray}%
where $c_{k},$ $k=1,2,3,4$ are arbitrary constants and $\partial
_{x}^{-1}=\int\limits_{x}^{\infty }dx$\ is indefinite integral with respect
to $x$.
\end{proposition}

\begin{proposition}
If the coefficients of (2.1) satisfies the properties $p_{1}=q_{1}$ and $%
p_{2}=q_{2}$ then the system of equations (2.4) becomes the following
negative order pair of equations:%
\begin{eqnarray}
p_{1tx} &=&p_{1}\left[ 2\partial _{x}^{-1}\left( p_{1}^{2}\right)
_{t}+\partial _{x}^{-1}\left( p_{2}^{2}\right) _{t}+c_{1}\right] +p_{2}\left[
\partial _{x}^{-1}\left( p_{1}p_{2}\right) _{t}+c_{2}\right] ,  \notag \\
&&  \TCItag{2.6} \\
p_{2tx} &=&p_{2}\left[ \partial _{x}^{-1}\left( p_{1}^{2}\right)
_{t}+2\partial _{x}^{-1}\left( p_{2}^{2}\right) _{t}+c_{3}\right] +p_{1}%
\left[ \partial _{x}^{-1}\left( p_{1}p_{2}\right) _{t}+c_{4}\right] ,  \notag
\end{eqnarray}%
where $c_{k},$ $k=1,2,3,4$ are arbitrary constants and $\partial
_{x}^{-1}=\int\limits_{x}^{\infty }dx$\ is indefinite integral with respect
to $x$.
\end{proposition}

\bigskip The proofs of these propositions are omitted since the proofs of
many more cases are provided in [15]. The following corollaries of
Propositions 1 and 2 are valid, respectively.

\begin{corollary}
(CKG Equation) In the case $p_{1}=p_{2}=p$ and $c_{1}=c_{2}=c_{3}=c_{4}=-%
\frac{1}{2}$ this nonlinear evolution equation (2.5) has the form 
\begin{equation}
p_{tx}=-4p\int\limits_{x}^{+\infty }\left( p^{2}\right) _{t}dx+p\text{.} 
\tag{2.7}
\end{equation}%
that becomes CKG by the substitution $p=\frac{ir}{\sqrt{2}}$ with real
valued $r(x)$, where the spectral problem is the \bigskip classical
Zakharov--Shabat system with real and anti-symmetric.
\end{corollary}

\begin{corollary}
\bigskip (nmKdV Equation) In the case $p_{1}=p_{2}=p$ and $%
c_{1}=c_{2}=c_{3}=c_{4}=\frac{1}{2}$ this nonlinear evolution equation (2.6)
has the form 
\begin{equation}
p_{tx}=4p\int\limits_{x}^{+\infty }\left( p^{2}\right) _{t}dx+p.  \tag{2.8}
\end{equation}%
that becomes nmKdV by the substitution $p=\frac{ir}{\sqrt{2}}$ with real
valued $r(x)$, where the spectral problem is the \bigskip classical
Zakharov--Shabat system with real and symmetric potential.
\end{corollary}

Really, the spectral problem for the equations (1.2$_{\pm }$) are the
\bigskip classical Zakharov--Shabat system in the following form:\textbf{\ }%
\begin{equation}
\left\{ 
\begin{array}{c}
u_{1x}=-i\mu u_{1}-ru_{2}, \\ 
u_{2x}=\pm ru_{1}+i\mu u_{2},%
\end{array}%
\right.  \tag{2.9}
\end{equation}%
that is produced from (1.2) if $p_{1}=p_{2}=\mp q_{1}=\mp q_{2}=p$ are taken
in (2.1), we obtain 
\begin{equation*}
\left\{ 
\begin{array}{c}
\varphi _{1x}=i\alpha _{1}\lambda \varphi _{1}+ip(\varphi _{2}+\varphi _{3}),
\\ 
\varphi _{2x}=\mp ip\varphi _{1}+i\alpha _{2}\lambda \varphi _{2}, \\ 
\varphi _{3x}=\mp ip\varphi _{1}+i\alpha _{2}\lambda \varphi _{3}.%
\end{array}%
\right.
\end{equation*}%
This system becomes

\begin{equation*}
\left\{ 
\begin{array}{c}
u_{1x}=i\alpha _{1}\lambda u_{1}+ipu_{2}, \\ 
u_{2x}=\mp 2ipu_{1}+i\alpha _{2}\lambda u_{2},%
\end{array}%
\right.
\end{equation*}%
by the substitutions $\sqrt{2}\varphi _{1}=u_{1},$ $\varphi _{2}+\varphi
_{3}=u_{2}$, $-\alpha _{1}=\alpha _{2}=\beta $, $\mu =\beta \lambda $ and $p=%
\frac{ir}{\sqrt{2}}$ transforms this system to classical ZS system in the
form (2.9).

\begin{remark}
\bigskip In the defocusing case when $q_{1}=p_{1}^{\ast }$ and $%
q_{2}=p_{2}^{\ast }$ the system (2.1) is called Manakov system [16], and the
case $q_{1}=-p_{1}^{\ast }$ and $q_{2}=-p_{2}^{\ast }$ makes the system
(2.1) focusing, [17,18]. Referring to the results for the inverse scattering
problem for the Manakov system (2.1) (see Appendix of the paper [15]), the
soliton solutions for the negative order AKNS equation (2.5) or (2.7) can be
examined similarly to the pair of nonlinear Schr\"{o}dinger equations, as
done in [15] for two-dimensional stationary self-focusing of electromagnetic
waves. However, of greater importance is the particular case of the
defocusing (or Manakov) system (2.1), where the nonlinear evolution equation
(2.5) when $p_{1}=p_{2}=p=\frac{ir}{\sqrt{2}}$ and $c_{1}=c_{2}=c_{3}=c_{4}=-%
\frac{1}{2}$ becomes the system which takes the form: 
\begin{equation*}
r_{tx}=2r\int\limits_{x}^{+\infty }\left( r^{2}\right) _{t}dx+r
\end{equation*}%
which is the nonlinear Klein--Gordon equation coupled with a scalar field
(1.2$_{+}$), where $r$ is real valued function. In the focusing case of
system (2.1), the evolution equation (2.6) when $p_{1}=p_{2}=p=\frac{ir}{%
\sqrt{2}}$ and $c_{1}=c_{2}=c_{3}=c_{4}=\frac{1}{2}$ becomes the form: 
\begin{equation*}
r_{tx}=-2r\int\limits_{x}^{+\infty }\left( r^{2}\right) _{t}dx+r
\end{equation*}%
which is the negative order mKdV equation, where $r$ is a real-valued
function.
\end{remark}

\section{Zakharov--Shabat System with Real Anti-Symmetric and Real Symmetric
Potential}

In this section, we recall the necessary results from $\left[ 19,20,21\right]
$\ on inverse scattering problem for classical Zakharov--Shabat system:%
\begin{equation*}
\left\{ 
\begin{array}{c}
u_{1x}=-i\mu u_{1}-ru_{2}, \\ 
u_{2x}=\pm ru_{1}+i\mu u_{2}%
\end{array}%
\right.
\end{equation*}%
with real coefficient $r=r(x)$. It can be becomes 
\begin{equation}
\left\{ 
\begin{array}{c}
u_{1x}=-i\mu u_{1}\mp ru_{2}, \\ 
u_{2x}=ru_{1}+i\mu u_{2}%
\end{array}%
\right.  \tag{3.1$_{\pm }$}
\end{equation}%
under the substitution $\pm u_{2}\rightarrow u_{2}$.

Consider more general system of two differential equations 
\begin{equation}
\left\{ 
\begin{array}{c}
u_{1x}=-i\mu u_{1}+qu_{2}, \\ 
u_{2x}=ru_{1}+i\mu u_{2}.%
\end{array}%
\right.   \tag{3.2}
\end{equation}%
Impose the condition on the functions $q(x)$ and $r(x)$ that decay
sufficiently rapidly as $\left\vert x\right\vert \rightarrow \infty $. This
system yields (3.1$_{\pm }$) if $q=\mp r$.

Let the eigenfunctions $\Phi ,\Psi ,$ $\overline{\Phi }$ and $\overline{\Psi 
}$ be defined with the following boundary conditions for the eigenvalue $\mu 
$ in ZS system

\begin{eqnarray*}
\Phi &\sim &\left( 
\begin{array}{c}
1 \\ 
0%
\end{array}%
\right) e^{-i\mu x},\text{ }\Psi \sim \left( 
\begin{array}{c}
0 \\ 
1%
\end{array}%
\right) e^{i\mu x}, \\
x &\rightarrow &-\infty \text{\ \ \ \ \ \ \ \ \ \ \ \ \ \ \ \ \ \ \ \ \ }%
x\rightarrow +\infty
\end{eqnarray*}%
\begin{eqnarray*}
\overline{\Phi } &\sim &\left( 
\begin{array}{c}
0 \\ 
-1%
\end{array}%
\right) e^{i\mu x},\text{ }\overline{\Psi }\sim \left( 
\begin{array}{c}
1 \\ 
0%
\end{array}%
\right) e^{-i\mu x}. \\
x &\rightarrow &-\infty \text{\ \ \ \ \ \ \ \ \ \ \ \ \ \ \ \ \ \ \ \ \ }%
x\rightarrow +\infty
\end{eqnarray*}

For these eigenfunctions, $W(\Phi ,\overline{\Phi })=-1$ and $W(\Psi ,%
\overline{\Psi })=-1,$ where $W$ is the Wronskian. Therefore, the
eigenfunctions $\Psi $ and $\overline{\Psi }$ are linearly independent.
Hence the functions $\Phi $ and $\overline{\Phi }$ can be written as

\begin{equation*}
\Phi =a(\mu )\overline{\Psi }+b(\mu )\Psi ,
\end{equation*}

\begin{equation*}
\overline{\Phi }=-\overline{a}(\mu )\Psi +\overline{b}(\mu )\overline{\Psi }.
\end{equation*}

The scattering matrix is usually defined as

\begin{equation*}
S=\left( 
\begin{array}{cc}
a & b \\ 
\overline{b} & -\overline{a}%
\end{array}%
\right) .
\end{equation*}

Using $W(\Phi ,\overline{\Phi })=-1$ equation,

\begin{equation*}
a(\mu )\overline{a}(\mu )+b(\mu )\overline{b}(\mu )=1,
\end{equation*}%
is obtained. The functions $e^{i\mu x}\Phi $, $e^{-i\mu x}\Psi $ admits
analytical continuation into upper half-plane of $\mu $ and $e^{-i\mu x}%
\overline{\Phi },$ $e^{i\mu x}\overline{\Psi }$ admits analytical
continuation into lower half-plane of $\mu .$ It follows from $a(\mu
)=W(\Phi ,\Psi )$ is analytic in the upper half-plane and $\overline{a}(\mu
)=W(\overline{\Phi },\overline{\Psi })$ is analytic in the lower half-plane;
moreover, they tends to unity as $\left\vert \mu \right\vert \rightarrow
\infty $.

The function $a(\mu )$ has a zero in the upper half plane and $\overline{a}%
(\mu )$ has a zero in the lower half plane. If the zeros of $a(\mu )$ are
called $\mu _{k},k=1,2,...,N$ then at $\mu =\mu _{k},$ $\Phi $ and $\Psi $
proportional such that

\begin{equation*}
\Phi =c_{k}\Psi .
\end{equation*}

Similarly, if the zeros of $\overline{a}(\mu )$ are called $\overline{\mu }%
_{k},k=1,2,...,\overline{N}$ then at $\mu =\overline{\mu }_{k},$ $\overline{%
\Phi }$ and $\overline{\Psi }$ proportional such that

\begin{equation*}
\overline{\Phi }=\overline{c_{k}}\overline{\Psi }.
\end{equation*}%
In this case, $b$ and $\overline{b}$ can be expanded to \ $c_{k}=b(\mu _{k})$
and $\overline{c_{k}}=\overline{b}(\overline{\mu }_{k}).$ In this case, $%
a(\mu )$ and $\overline{a}(\mu )$ are analytic on the real axis, and are
also analytic in the upper half plane and the lower half plane. This means
that $a(\mu )$ has only a finite number of zeros for $\func{Im}(\mu )\geqq
0. $

The special type of relationship between $r$ and $q$, i.e., $q=\mp r$, a
case of special interest. If, in ZS system, we put 
\begin{equation*}
q=\mp r
\end{equation*}%
then involution arises in the solutions of ZS system. In other words, if $%
\Phi ==\left( 
\begin{array}{c}
\Phi _{1} \\ 
\Phi _{2}%
\end{array}%
\right) $ and $\Psi ==\left( 
\begin{array}{c}
\Psi _{1} \\ 
\Psi _{2}%
\end{array}%
\right) $ are the solutions of ZS system with real $\mu $ then the columns 
\begin{equation*}
\overline{\Phi }(x,\mu )=\left( 
\begin{array}{c}
\Phi _{2}(x,-\mu ) \\ 
\mp \Phi _{1}(x,-\mu )%
\end{array}%
\right) \text{ and }\overline{\Psi }\text{ }(x,\mu )=\left( 
\begin{array}{c}
\pm \Psi _{2}(x,-\mu ) \\ 
-\Psi _{1}(x,-\mu )%
\end{array}%
\right)
\end{equation*}%
also are solutions and implies that symmetry relations

\begin{equation*}
\overline{a}(\mu )=a(-\mu ),\text{ }\overline{b}(\mu )=\pm b(-\mu )
\end{equation*}%
and 
\begin{equation*}
\overline{N}=N,\text{ }\overline{\mu _{k}}=-\mu _{k},\text{ }\overline{c_{k}}%
=\pm c_{k}.
\end{equation*}%
Hence if $q=\mp r$ and $r$ is real then all of above symmetry conditions
hold. This implies that when $\mu _{k}$ is eigenvalue then so is $-\mu
_{k}^{\ast }$. This means eigenvalues either are on the imaginary axis or
are paired.

Now let's examine the inverse scattering problem. \ 

Let the $\Psi $ and $\overline{\Psi }$ functions be expressed with the
integral representations

\begin{equation*}
\Psi =\left( 
\begin{array}{c}
0 \\ 
1%
\end{array}%
\right) e^{i\mu x}+\underset{x}{\overset{+\infty }{\int }}K(x,s)e^{i\mu s}ds,
\end{equation*}

\begin{equation*}
\overline{\Psi }=\left( 
\begin{array}{c}
1 \\ 
0%
\end{array}%
\right) e^{-i\mu x}+\underset{x}{\overset{+\infty }{\int }}\overline{K}%
(x,s)e^{-i\mu s}ds,
\end{equation*}%
where $K(x,s)=\left( 
\begin{array}{c}
K_{1}(x,s) \\ 
K_{2}(x,s)%
\end{array}%
\right) ,$ $\overline{K}(x,s)=\left( 
\begin{array}{c}
\overline{K_{1}}(x,s) \\ 
\overline{K_{2}}(x,s)%
\end{array}%
\right) .$

It is necessary and sufficient to have

\begin{equation*}
\left( \partial _{x}-\partial _{s}\right) K_{1}(x,s)-q(x)K_{2}(x,s)=0,
\end{equation*}

\begin{equation*}
\left( \partial _{x}+\partial _{s}\right) K_{2}(x,s)-r(x)K_{1}(x,s)=0,
\end{equation*}%
subject to the boundary conditions

\begin{equation*}
K_{1}(x,x)=-\frac{1}{2}q(x),\text{ }\underset{s\rightarrow \infty }{\lim }%
K(x,s)=0.
\end{equation*}%
These equations can be expressed with matrices

\begin{equation*}
\widetilde{K}=\left( 
\begin{array}{cc}
\overline{K_{1}} & K_{1} \\ 
\overline{K_{2}} & K_{2}%
\end{array}%
\right) ,\widetilde{F}=\left( 
\begin{array}{cc}
0 & -\overline{F} \\ 
F & 0%
\end{array}%
\right) ,
\end{equation*}%
where by we have Gel'fand-Levitan-Marchenko (GLM) equation

\begin{equation*}
\widetilde{K}(x,y)+\widetilde{F}(x+y)+\underset{x}{\overset{+\infty }{\int }}%
\widetilde{K}(x,s)\widetilde{F}(s+y)ds=0,
\end{equation*}%
where 
\begin{eqnarray*}
F(x) &=&\frac{1}{2\pi }\underset{-\infty }{\overset{+\infty }{\int }}\frac{%
b(\mu )}{a(\mu )}e^{i\mu x}d\mu -i\overset{N}{\underset{j=1}{\sum }}%
c_{j}e^{i\mu _{j}x}, \\
\text{ }\overline{F}(x) &=&\frac{1}{2\pi }\underset{-\infty }{\overset{%
+\infty }{\int }}\frac{\bar{b}(\mu )}{\bar{a}(\mu )}e^{-i\mu x}d\mu -i%
\overset{N}{\underset{j=1}{\sum }}\bar{c}_{j}e^{-i\overline{\mu _{j}}x}
\end{eqnarray*}

With the symmetry conditions (when $q=\mp r$ and $r$ is real, then $F$ and $%
K(x,z)$ are real)

\begin{equation*}
\overline{F}(x)=F(x),\text{ }\overline{K}(x,y)=\left( 
\begin{array}{c}
K_{2}(x,y) \\ 
-K_{1}(x,y)%
\end{array}%
\right)
\end{equation*}%
the GLM equation becomes

\begin{equation*}
K_{1}\left( x,y\right) \pm F\left( x+y\right) \mp \underset{x}{\overset{%
+\infty }{\int }}\underset{x}{\overset{+\infty }{\int }}K_{1}\left(
x,z\right) F\left( z+s\right) F\left( s+y\right) dsdz=0
\end{equation*}%
with the reconstruction formula

\begin{equation*}
r(x)=\pm 2K_{1}(x,x).
\end{equation*}

The inverse problem equation show that the potential in ZS system is
uniquely defined by the set 
\begin{equation*}
S=\left\{ \text{ }\mu _{n},\text{ }c_{n}\text{, }n=1,2,...,N;\text{ }\frac{%
a(\mu )}{b(\mu )},\text{ }-\infty <\mu <+\infty \right\}
\end{equation*}%
which is called scattering data for ZS equation.

If $r(x)$ is finite function, the solutions $\Phi $ and $\Psi $ are entire
functions of $\mu $, and $b(\mu )$ may be well-defined over the complex
plane. And, if $\mu _{k}$ is a zero of $a(\mu )$, the normalization factor $%
c_{k}$ given by $\Phi \left( x,\mu _{k}\right) =c_{k}\Psi \left( x,\mu
_{k}\right) $ is 
\begin{equation}
c_{k}=b(\mu _{k})\text{.}  \tag{3.3}
\end{equation}%
Thus, the symmetry conditions implies that if $\mu _{k}$ lies on the
imaginary axis, then $\func{Re}c_{k}=0$.

\section{N-Soliton Solutions of Coupled Klein - Gordon\ and Negative Order
mKdV Equations}

The N-soliton solutions of equation (1.2)\ (equation (2.1$_{+}$) or equation
(2.1$_{-}$)) will be studied by using the inverse scattering method. The
Gel'fand--Levitan--Marchenko (GLM) equation corresponding to the
Zakharov-Shabat system (3.1) (system (3.1$_{+}$) or system (3.1$_{-}$)) will
be applied, respectively.

It is our close-up aim to study evolution of scattering data $\left\{ \frac{%
a(\mu )}{b(\mu )};\text{ }\mu _{n},\text{ }c_{n}\text{, }n=1,2,...,N\right\} 
$ for the ZS system (3.1) when the potential $r(x,t)$ satisfies the CKG or
nmKdV equations, respectively.

As is shown in previous section, the CKG equation admits the Lax
representation which the components are ZS system with real and
anti-symmetric potential and the another component is the form:%
\begin{equation}
\left[ 
\begin{array}{c}
u_{1t} \\ 
u_{2t}%
\end{array}%
\right] =\left[ 
\begin{array}{cc}
\frac{A(x,t)}{\mu } & \frac{B(x,t)}{\mu } \\ 
\frac{C(x,t)}{\mu } & -\frac{A(x,t)}{\mu }%
\end{array}%
\right] \left[ 
\begin{array}{c}
u_{1} \\ 
u_{2}%
\end{array}%
\right] ,  \tag{4.1}
\end{equation}%
where the compatibility condition becomes

\begin{eqnarray*}
A_{x} &=&-r(C+B), \\
B_{x} &=&2Ar,\text{ }C_{x}=2Ar, \\
r_{t} &=&-2iB,\text{ }r_{t}=-2iC.
\end{eqnarray*}

It is clear to see that this system produce the equation 
\begin{equation*}
r_{tx}=2r\int\limits_{x}^{+\infty }\left( r^{2}\right) _{t}dx-4iC_{0}r
\end{equation*}%
which is CKG equation when $C_{0}=\frac{i}{4}$. This constant appears in
finding of $A$ in the following form 
\begin{equation*}
A=i\int\limits_{x}^{+\infty }rr_{t}dx+C_{0}
\end{equation*}%
that is important for the evolution of reflection coefficient $b(\mu )$.

\bigskip Same is shown for nmKdV equation that admits the Lax representation
which the components are ZS system with real and symmetric potential and the
another component is the form (4.1), where the compatibility condition
becomes

\begin{eqnarray*}
A_{x} &=&r(C-B), \\
B_{x} &=&-2Ar,\text{ }C_{x}=2Ar, \\
r_{t} &=&2iB,\text{ }r_{t}=-2iC.
\end{eqnarray*}

It is clear to see that this system produce the equation 
\begin{equation*}
r_{tx}=-2r\int\limits_{x}^{+\infty }\left( r^{2}\right) _{t}dx-4iC_{1}r
\end{equation*}%
which is nmKdV equation when $C_{1}=\frac{i}{4}$. This constant appears in
finding of $A$ in the following form 
\begin{equation*}
A=-i\int\limits_{x}^{+\infty }rr_{t}dx+C_{1}
\end{equation*}%
that is important for the evolution of reflection coefficient $b(\mu )$.

\begin{theorem}
Let $r(x,t)$ be a coefficient of the system (3.1) satisfying the equation
(1.2), then the evolution of the scattering data of this system (3.1) is the
following form:

\begin{equation*}
a(\mu ,t)=a(\mu ,0),
\end{equation*}%
\begin{equation}
b(\mu ,t)=b(\mu ,0)e^{-\frac{i}{2\mu }t},  \tag{4.2}
\end{equation}%
\begin{equation*}
\mu _{j}(t)=\mu _{j};\text{ }c_{n}(t)=c_{j,0}e^{-\frac{i}{2\mu _{_{j}}}t},
\end{equation*}

where $c_{j,0}=c_{j}(t=0).$

\begin{proof}
Requiring that $r\rightarrow 0$ as $\left\vert x\right\vert \rightarrow
+\infty $ gives us a large class of equations with the property that $%
A\rightarrow A\_(\mu ),$ $D\rightarrow -A\_(\mu ),$ $B,$ $C\rightarrow 0$
for $x\rightarrow +\infty $. The time dependence of the eigenfunctions $\Phi
,\Psi ,\bar{\Phi}$ and $\overline{\Psi }$ are defined in the following form:

\begin{eqnarray*}
\Phi ^{(t)} &=&\Phi e^{A\_(t)},\text{ }\Psi ^{(t)}=\Psi e^{-A\_(t)}, \\
\overline{\Phi }^{(t)} &=&\overline{\Phi }e^{-A\_(t)},\text{ }\overline{\Psi 
}^{(t)}=\overline{\Psi }e^{A\_(t)}.
\end{eqnarray*}

The time evolution of $\Phi ^{(t)}$

\begin{equation*}
\frac{d\Phi ^{(t)}}{dt}=\left( 
\begin{array}{cc}
\frac{A}{\mu } & \frac{B}{\mu } \\ 
\frac{C}{\mu } & -\frac{A}{\mu }%
\end{array}%
\right) \Phi ^{(t)},
\end{equation*}

shows that $\Phi $ satisfies the equation

\begin{equation*}
\frac{d\Phi }{dt}=\left( 
\begin{array}{cc}
\frac{A}{\mu }-A\_(\mu ) & \frac{B}{\mu } \\ 
\frac{C}{\mu } & -\frac{A}{\mu }-A\_(\mu )%
\end{array}%
\right) \Phi .
\end{equation*}

If we use

\begin{equation*}
\Phi =a\overline{\Psi }+b\Psi \sim a\left( 
\begin{array}{c}
1 \\ 
0%
\end{array}%
\right) e^{-i\mu x}+b\left( 
\begin{array}{c}
0 \\ 
1%
\end{array}%
\right) e^{i\mu x}\text{ }(x\sim \infty )
\end{equation*}

then the last two relations yields

\begin{equation*}
\left( 
\begin{array}{c}
a_{t}e^{-i\mu x} \\ 
b_{t}e^{i\mu x}%
\end{array}%
\right) =\left( 
\begin{array}{c}
0 \\ 
-2A\_(\mu )be^{i\mu x}%
\end{array}%
\right) .
\end{equation*}%
as $x\rightarrow +\infty $.

Therefore, the following equations are obtained:%
\begin{equation*}
a_{t}=0,\text{ }b_{t}=-2A\_(\mu )b
\end{equation*}%
with $A_{-}(\mu )=\frac{C_{0}}{\mu }=\frac{i}{4\mu }$. From this equations
we find that

\begin{equation*}
a(\mu ,t)=a(\mu ,0),\text{ }b(\mu ,t)=b(\mu ,0)e^{-\frac{i}{2\mu }t}\text{. }
\end{equation*}

The eigenvalues $\mu _{j}$ are, being the zeros of time-invariant function $%
a(\mu )$, also time-invariant $\mu _{j}(t)=\mu _{j}$. The time-dependence of
the normalization factors $c_{n}$ can be found directly from (3.3): 
\begin{equation*}
c_{j}(t)=c_{j,0}e^{-\frac{i}{2\mu _{_{j}}}t},j=1,2,...,N.
\end{equation*}
\end{proof}
\end{theorem}

Let $r(x,t)$ be a coefficient of the ZS system with real and anti-symmetric
potential (system (3.1$_{+}$)) which satisfies the CKG equation (1.2$_{+}$)
or a coefficient of the ZS system with real and symmetric potential (system
(3.1$_{-}$)) which satisfies the nmKdV equation (1.2$_{-}$), then the
corresponding GLM equation for this systems are as follows

\begin{equation}
K\left( x,y;t\right) \pm F\left( x+y;t\right) \mp \underset{x}{\overset{%
+\infty }{\int }}\underset{x}{\overset{+\infty }{\int }}K\left( x,z;t\right)
F\left( z+s;t\right) F\left( s+y;t\right) dsdz=0,  \tag{4.3}
\end{equation}%
where

\begin{equation}
F\left( s;t\right) =\frac{1}{2\pi }\underset{-\infty }{\overset{+\infty }{%
\int }}\frac{b(\mu ,0)}{a(\mu ,0)}e^{i\mu s-\frac{i}{2\mu }t}d\mu -i\overset{%
N}{\underset{j=1}{\sum }}c_{j,0}e^{i\mu _{j}s-\frac{i}{2\mu _{j}}t}. 
\tag{4.4}
\end{equation}%
Also, the coefficient system (3.1$_{\pm }$) becomes

\begin{equation*}
r(x,t)=\pm 2K(x,x;t).
\end{equation*}%
In the case when the reflection coefficient disappear in (4.4) (or, system
(3.1$_{\pm }$) has only discrete spectrum) then the equation (4.3) is the
integral equation with degenerate kernel.

The following theorem is true for the N-soliton solutions for (1.2$_{\pm }$).

\begin{theorem}
Let $r(x,t)$ be a coefficient of the system (3.1$_{\pm }$) satisfying the
equation (1.2$_{\pm }$), and also system (3.1$_{\pm }$) has only discrete
spectrum on the imaginary axis. Then the N-soliton solution of (1.2$_{\pm }$%
) is obtained as a solution of the inverse problem in the following form:

\begin{equation}
r(x,t)=2\overset{N}{\underset{j,m=1}{\sum }}\frac{c_{j,0}c_{m,0}}{i\left(
\mu _{m}+\mu _{j}\right) }e^{i\left( 2\mu _{m}+\mu _{j}\right) x}e^{-i\left( 
\frac{1}{2\mu _{_{m}}}+\frac{1}{2\mu _{j}}\right) t}\chi _{j}(x,t)+2i\overset%
{N}{\underset{j=1}{\sum }}c_{j,0}e^{2i\mu _{j}x-\frac{i}{2\mu _{j}}t}, 
\tag{4.5}
\end{equation}

where $\chi _{1},\chi _{2},...,\chi _{N}$ are the column entries of 
\begin{equation*}
\chi =\mp (I\mp \Lambda ^{2})^{-1}\Lambda e^{i\mu x}.
\end{equation*}
Here $\Lambda $ is $N\times N$ matrix whose components are%
\begin{equation*}
\Lambda _{nm}=\frac{c_{m,0}}{\mu _{m}+\mu _{n}}e^{i\left( \mu _{m}+\mu
_{n}\right) x-\frac{i}{2\mu _{_{m}}}t},
\end{equation*}

$e^{i\mu x}$ is the columns vector with the entries $e^{i\mu _{1}x},e^{i\mu
_{2}x},...,e^{i\mu _{N}x}$ and $\mu _{n}$ are eigenvalues, $c_{n,0}$ are
normalized numbers of the ZS system (3.1$_{\pm }$).

\begin{proof}
Let $r(x,t)$ be a coefficient of the ZS system with real and anti-symmetric
potential (system (3.1$_{+}$)) which satisfies the CKG equation (1.2$_{+}$)
or a coefficient of the ZS system with real and symmetric potential (system
(3.1$_{-}$)) which satisfies the nmKdV equation (1.2$_{-}$) and also these
systems have only discrete spectrum. The equation (4.3) is a integral
equation with degenered kernel for arbitrary fixed $x$ and $t$:%
\begin{eqnarray*}
&&K\left( x,y;t\right) \pm \underset{x}{\overset{+\infty }{\int }}\underset{x%
}{\overset{+\infty }{\int }}K\left( x,z;t\right) \overset{N}{\underset{j,m=1}%
{\sum }}c_{j,0}c_{m,0}e^{i\left( \mu _{m}+\mu _{j}\right) s}e^{i\mu _{m}y-%
\frac{i}{2\mu _{_{m}}}t}e^{i\mu _{j}z-\frac{i}{2\mu _{j}}t}dsdz \\
&=&\pm i\overset{N}{\underset{j=1}{\sum }}c_{j,0}e^{i\mu _{j}\left(
x+y\right) -\frac{i}{2\mu _{j}}t},
\end{eqnarray*}%
The solution of (4.3) in the following form:

\begin{equation}
K\left( x,y;t\right) \mp \overset{N}{\underset{j,m=1}{\sum }}\frac{%
c_{j,0}c_{m,0}}{i\left( \mu _{m}+\mu _{j}\right) }e^{i\left( \mu _{m}+\mu
_{j}\right) x}e^{i\mu _{m}y-\frac{i}{2\mu _{_{m}}}t}e^{-\frac{i}{2\mu _{j}}%
t}\chi _{j}(x,t)=\pm i\overset{N}{\underset{j=1}{\sum }}c_{j,0}e^{i\mu
_{j}\left( x+y\right) -\frac{i}{2\mu _{j}}t},  \tag{4.6}
\end{equation}

where is chosen $\chi _{j}(x,t)=\overset{+\infty }{\int_{x}}K(x,z;t)e^{i\mu
_{j}z}dz$ and is the solution of following system of equations:

\begin{eqnarray*}
&&\chi _{n}(x,t)\mp \overset{N}{\underset{j,m=1}{\sum }}\frac{c_{j,0}}{\mu
_{m}+\mu _{j}}e^{i\left( \mu _{m}+\mu _{j}\right) x-\frac{i}{2\mu _{j}}t}%
\frac{c_{m,0}}{\mu _{m}+\mu _{n}}e^{i\left( \mu _{m}+\mu _{n}\right) x-\frac{%
i}{2\mu _{_{m}}}t}\chi _{j}(x,t) \\
&=&\overset{N}{\underset{j=1}{\mp \sum }}\frac{c_{j,0}}{\mu _{j}+\mu _{n}}%
e^{i\left( \mu _{j}+\mu _{n}\right) x}e^{i\mu _{j}x-\frac{i}{2\mu _{j}}t},%
\text{ }n=1,2,...,N.
\end{eqnarray*}

Introducing the matrix $\Lambda $ with the entries $\Lambda _{nm}=\frac{%
c_{m,0}}{\mu _{m}+\mu _{n}}e^{i\left( \mu _{m}+\mu _{n}\right) x-\frac{i}{%
2\mu _{_{m}}}t}$ and denoting by $\chi $ and $e^{i\mu x}$ the columns $\chi
_{1},\chi _{2},...,\chi _{N}$ and $e^{i\mu _{1}x},e^{i\mu _{2}x},...,e^{i\mu
_{N}x}$ this equation can be written in the following form

\begin{equation*}
(I\mp \Lambda ^{2})\chi =\mp \Lambda e^{i\mu x},
\end{equation*}

where $I$ is the $N\times N$ identity matrix.

Thus, we have 
\begin{equation*}
\chi =\mp (I\mp \Lambda ^{2})^{-1}\Lambda e^{i\mu x}.
\end{equation*}

Due to $r(x,t)=\pm 2K(x,x;t)$, from (4.6) the coefficient $r(x,t)$ is found
as equation (4.5).
\end{proof}
\end{theorem}

\bigskip If $\mu _{j}=i\kappa _{j},$ $\kappa _{j}>0$ is chosen in system
(4.2) then $\func{Re}c_{j,0}=0$ and the more suitable formula

\begin{equation*}
c_{j}(t)=i\omega _{j}e^{-\frac{t}{2\kappa _{j}}}
\end{equation*}%
are obtained, where $i\omega _{j}=c_{j,0}$.\bigskip 

If scattering data (4.2) is substituted in equation (4.4), and let $\omega
_{j}=-ic_{j,0}$ \ and $\mu _{j}=i\kappa _{j}$ are chosen. If the system (3.1$%
_{\pm }$) has only discrete spectrum the formula (4.4) becomes 
\begin{equation*}
F(x+y;t)=\overset{N}{\underset{j=1}{\sum }}\omega _{j}e^{-\kappa _{j}(x+y)-%
\frac{t}{2\kappa _{_{j}}}}
\end{equation*}

By adding it in (4.3), the following equation%
\begin{eqnarray*}
&&K(x,y;t)\mp \underset{x}{\overset{+\infty }{\int }}\underset{x}{\overset{%
+\infty }{\int }}K(x,z;t)\overset{N}{\underset{j,m=1}{\sum }}\omega
_{j}\omega _{m}e^{-\kappa _{j}(z+s)-\frac{t}{2\kappa _{j}}}e^{-\kappa
_{m}(s+y)-\frac{t}{2\kappa _{_{m}}}}dsdz \\
&=&\mp \overset{N}{\underset{j=1}{\sum }}\omega _{j}e^{-\kappa _{j}(x+y)-%
\frac{t}{2\kappa _{j}}}
\end{eqnarray*}%
is obtained. From this equation

\begin{eqnarray}
&&K\left( x,y;t\right) \mp \overset{N}{\underset{j,m=1}{\sum }}\frac{\omega
_{j}\omega _{m}}{\kappa _{j}+\kappa _{m}}e^{-\left( \kappa _{j}+\kappa
_{m}\right) x-\frac{t}{2\kappa _{m}}-\frac{t}{2\kappa _{j}}}e^{-\kappa
_{m}y}\left( \overset{+\infty }{\int_{x}}K(x,z;t)e^{-\kappa _{j}z}dz\right) 
\notag \\
&=&\overset{N}{\underset{j=1}{\mp \sum }}\omega _{j}e^{-\kappa _{j}(x+y)-%
\frac{t}{2\kappa _{j}}}  \TCItag{4.7}
\end{eqnarray}%
is obtained the following corollary of Theorem 2.

\begin{corollary}
Let $r(x,t)$ be a coefficient of the system (3.1$_{\pm }$) satisfying the
equation (1.2$_{\pm }$), and also system (3.1$_{\pm }$) has only discrete
spectrum on the imaginary axis. Then the solution of the inverse problem for
this system is%
\begin{equation}
r(x,t)=2\overset{N}{\underset{j,m=1}{\sum }}\frac{\omega _{j}\omega _{m}}{%
\kappa _{j}+\kappa _{m}}e^{-\left( 2\kappa _{j}+\kappa _{m}\right) x-\frac{t%
}{2\kappa _{m}}-\frac{t}{2\kappa _{j}}}\chi _{j}(x,t)-2\overset{N}{\underset{%
j=1}{\sum }}\omega _{j}e^{-2\kappa _{j}x-\frac{t}{2\kappa _{j}}},  \tag{4.8}
\end{equation}%
where $\chi _{1},\chi _{2},...,\chi _{N}$ are the column entries of $\chi
=\mp (I\mp \Pi ^{2})^{-1}\Pi e^{-\kappa x}.$ Here $\Lambda $ is $N\times N$
matrix whose components are%
\begin{equation*}
\Pi _{nm}=\frac{\omega _{m}}{\kappa _{n}+\kappa _{m}}e^{-\left( \kappa
_{n}+\kappa _{m}\right) x-\frac{t}{2\kappa _{m}}},
\end{equation*}%
$e^{-\kappa x}$ is the columns vector with the entries $e^{-\kappa
_{1}x},e^{-\kappa _{2}x},...,e^{-\kappa _{N}x}$ and $\mu _{n}=i\kappa _{n}$
are eigenvalues, $c_{n,0}=i\omega _{n}$ are normalized numbers of the ZS
system (3.1$_{\pm }$).
\end{corollary}

\textbf{Examples:} Now we shall consider certain simple cases. 

\textbf{a)} Let $a(\mu )$ have only one zero on the imaginary axis at the
point $\mu _{1}=i\kappa _{1}$ with the normalizing factor $c_{0}=i\omega _{1}
$ where $\kappa _{1}>0$ and $\omega _{1}$ should necessarily be real.

In the case $q=r$ the GLM equation has the form

\begin{equation*}
K(x,y;t)+\omega _{1}^{2}\underset{x}{\overset{+\infty }{\int }}\underset{x}{%
\overset{+\infty }{\int }}e^{-\kappa _{1}\left( z+2s+y\right) -\frac{t}{%
\kappa _{1}}}K(x,z;t)dsdz=\omega _{1}e^{-\kappa _{1}\left( x+y\right) -\frac{%
t}{2\kappa _{1}}}
\end{equation*}

The solution of this equation is

\begin{equation*}
K(x,y;t)=\frac{4\omega _{1}\kappa _{1}^{2}e^{-\kappa _{1}(x+y)-\frac{t}{%
2\kappa _{1}}}}{4\kappa _{1}^{2}+\omega _{1}^{2}e^{-4\kappa _{1}x-\frac{t}{%
\kappa _{1}}}}
\end{equation*}%
and the solution of GLM is

\begin{equation*}
r(x,t)=-\frac{8\omega _{1}\kappa _{1}^{2}}{4\kappa _{1}^{2}e^{\theta
}+\omega _{1}^{2}e^{-\theta }},\text{ \ }\theta =2\kappa _{1}x+\frac{t}{%
2\kappa _{1}}
\end{equation*}%
since $r(x,t)=2K(x,x;t)$.

In the case $q=-r$ the GLM equation is

\begin{equation*}
K(x,y;t)-\underset{x}{\overset{+\infty }{\omega _{1}^{2}\int }}\underset{x}{%
\overset{+\infty }{\int }}e^{-\kappa _{1}\left( z+2s+y\right) -\frac{t}{%
\kappa _{1}}}K(x,z;t)dsdz=-\omega _{1}e^{-\kappa _{1}\left( x+y\right) -%
\frac{t}{2\kappa _{1}}}.
\end{equation*}%
The solution of this equation is 
\begin{equation*}
K(x,y;t)=-\frac{4\omega _{1}\kappa _{1}^{2}e^{-\kappa _{1}(x+y)-\frac{t}{%
2\kappa _{1}}}}{4\kappa _{1}^{2}-\omega _{1}^{2}e^{-4\kappa _{1}x-\frac{t}{%
\kappa _{1}}}}
\end{equation*}%
and the solution of nmKdV equation is

\begin{equation*}
r(x,t)=-\frac{8\omega _{1}\kappa _{1}^{2}}{4\kappa _{1}^{2}e^{\theta
}-\omega _{1}^{2}e^{-\theta }},\text{ \ }\theta =2\kappa _{1}x+\frac{t}{%
2\kappa _{1}}
\end{equation*}%
since $r(x,t)=-2K(x,x;t)$.

These solutions are the single solitons of CKG and nmKdV moving a velocity $%
\frac{1}{4\kappa _{1}^{2}}$.

\textbf{b)} The general two-soliton solution of (1.2$_{\pm }$) is obtained
from (4.8), when $a(\mu )$ has two zeros on the imaginary axis at the points 
$\mu _{1}=i\kappa _{1}$ and $\mu _{2}=i\kappa _{2}$ ($\kappa _{1},\kappa
_{2}>0$) with the normalizing factor $c_{1}=i\omega _{1}$ and $c_{2}=i\omega
_{2}$: 
\begin{eqnarray*}
r(x,t) &=&\chi _{1}(x,t)\left( \pm \frac{\omega _{1}^{2}}{\kappa _{1}}%
e^{-3\kappa _{1}x-\frac{t}{\kappa _{1}}}\pm \frac{2\omega _{1}\omega _{2}}{%
\kappa _{1}+\kappa _{2}}e^{-\left( 2\kappa _{2}+\kappa _{1}\right) x-\frac{t%
}{2\kappa _{2}}-\frac{t}{2\kappa _{1}}}\right)  \\
&&+\chi _{2}(x,t)\left( \pm \frac{2\omega _{2}\omega _{1}}{\kappa
_{1}+\kappa _{2}}e^{-\left( 2\kappa _{1}+\kappa _{2}\right) x-\frac{t}{%
2\kappa _{1}}-\frac{t}{2\kappa _{2}}}\pm \frac{\omega _{2}^{2}}{\kappa _{2}}%
e^{-3\kappa _{2}x-\frac{t}{\kappa _{2}}}\right)  \\
&&\mp 2\omega _{1}e^{-2\kappa _{1}x-\frac{t}{2\kappa _{j}}}\mp 2\omega
_{2}e^{-2\kappa _{2}x-\frac{t}{2\kappa _{2}}},
\end{eqnarray*}%
where 
\begin{equation*}
\chi _{1}(x,t)=\frac{\mp \frac{\omega _{1}}{2\kappa _{1}}%
X_{1}X_{2}^{4}T_{1}T_{2}^{2}\mp \frac{\omega _{2}}{\kappa _{1}+\kappa _{2}}%
X_{1}^{3}X_{2}^{2}T_{1}^{2}T_{2}+\frac{\omega _{1}\omega _{2}^{2}(\kappa
_{1}-\kappa _{2})^{3}}{8\kappa _{1}\kappa _{2}^{2}(\kappa _{1}+\kappa
_{2})^{3}}X_{1}T_{1}}{X_{1}^{4}X_{2}^{4}T_{1}^{2}T_{2}^{2}\mp \frac{2\omega
_{1}\omega _{2}}{(\kappa _{1}+\kappa _{2})^{2}}X_{1}^{2}X_{2}^{2}T_{1}T_{2}%
\mp \frac{\omega _{1}^{2}}{4\kappa _{1}^{2}}X_{2}^{4}T_{2}^{2}\mp \frac{%
\omega _{2}^{2}}{4\kappa _{2}^{2}}X_{1}^{4}T_{1}^{2}+\frac{\omega
_{1}^{2}\omega _{2}^{2}(\kappa _{1}-\kappa _{2})^{4}}{16\kappa
_{1}^{2}\kappa _{2}^{2}(\kappa _{1}+\kappa _{2})^{4}}}
\end{equation*}%
and

\begin{equation*}
\chi _{2}(x,t)=\frac{\mp \frac{\omega _{1}}{\kappa _{1}+\kappa _{2}}%
X_{1}^{2}X_{2}^{3}T_{1}T_{2}^{2}\mp \frac{\omega _{2}}{2\kappa _{2}}%
X_{1}^{4}X_{2}T_{1}^{2}T_{2}-\frac{\omega _{1}^{2}\omega _{2}(\kappa
_{1}-\kappa _{2})^{3}}{8\kappa _{1}^{2}\kappa _{2}(\kappa _{1}+\kappa
_{2})^{3}}X_{2}T_{2}}{X_{1}^{4}X_{2}^{4}T_{1}^{2}T_{2}^{2}\mp \frac{2\omega
_{1}\omega _{2}}{(\kappa _{1}+\kappa _{2})^{2}}X_{1}^{2}X_{2}^{2}T_{1}T_{2}%
\mp \frac{\omega _{1}^{2}}{4\kappa _{1}^{2}}X_{2}^{4}T_{2}^{2}\mp \frac{%
\omega _{2}^{2}}{4\kappa _{2}^{2}}X_{1}^{4}T_{1}^{2}+\frac{\omega
_{1}^{2}\omega _{2}^{2}(\kappa _{1}-\kappa _{2})^{4}}{16\kappa
_{1}^{2}\kappa _{2}^{2}(\kappa _{1}+\kappa _{2})^{4}}}
\end{equation*}%
with the denotings 
\begin{equation*}
X_{1}=e^{\kappa _{1}x},\text{ }X_{2}=e^{\kappa _{2}x},\text{ }T_{1}=e^{\frac{%
t}{2\kappa _{1}}},\text{ }T_{2}=e^{\frac{t}{2\kappa _{2}}}\text{,}
\end{equation*}

\bigskip In result the general two-soliton solution is

\begin{equation*}
r(x,t)=\frac{\pm \frac{\omega _{1}^{2}\omega _{2}(\kappa _{1}-\kappa
_{2})^{2}}{2\kappa _{1}^{2}(\kappa _{1}+\kappa _{2})^{2}}X_{2}^{2}T_{2}\pm 
\frac{\omega _{1}\omega _{2}^{2}(\kappa _{1}-\kappa _{2})^{2}}{2\kappa
_{2}^{2}(\kappa _{1}+\kappa _{2})^{2}}X_{1}^{2}T_{1}-2\omega
_{1}X_{1}^{2}X_{2}^{4}T_{1}T_{2}^{2}-2\omega
_{2}X_{1}^{4}X_{2}^{2}T_{1}^{2}T_{2}}{X_{1}^{4}X_{2}^{4}T_{1}^{2}T_{2}^{2}%
\mp \frac{2\omega _{1}\omega _{2}}{(\kappa _{1}+\kappa _{2})^{2}}%
X_{1}^{2}X_{2}^{4}T_{1}T_{2}\mp \frac{\omega _{1}^{2}}{4\kappa _{1}^{2}}%
X_{2}^{4}T_{2}^{2}\mp \frac{\omega _{2}^{2}}{4\kappa _{2}^{2}}%
X_{1}^{4}T_{1}^{2}+\frac{\omega _{1}^{2}\omega _{2}^{2}(\kappa _{1}-\kappa
_{2})^{4}}{16\kappa _{1}^{2}\kappa _{2}^{2}(\kappa _{1}+\kappa _{2})^{4}}}%
\text{.}
\end{equation*}

Another elementary formation is the solution of (4.5) (this solution is
called twin or pulsing soliton) when $a(\mu )$ has a pair of zeros
symmetrically located relative to the imaginary axis: $\mu _{1}=\theta
+i\sigma $ and $\mu _{2}=-\theta +i\sigma $ with $c_{1,0}=c_{2,0}=c$.

\section{Conclusion}

In this paper, we study the soliton solutions of the Klein-Gordon equation
coupled with a scalar field (CKG), which shares the same bilinear form with
the sine-Gordon equation and the negative order modified Korteweg-de Vries
(nmKdV) equation. We found the Lax pair of the CKG and nmKdV equations as
the negative order AKNS type. The spectral problem is the ZS system with
real, anti-symmetric, and real symmetric potential. This makes it possible
to use the inverse scattering method's technique via the linear GLM integral
equation to obtain and analyze the soliton solutions of the CKG and nmKdV.
This method allows us to demonstrate the complete integrability of the
coupled Klein-Gordon and negative order modified Korteweg-de Vries
equations. On the other hand, while various extensions and generalizations
of the inverse scattering method have been discovered, it seems that many
different integrable negative order nonlinear wave equations still remain to
be found.

\end{document}